\documentstyle[twocolumn,aps,prl]{revtex}

\begin{document}
\draft
\title{ Localization Transition  in Incommensurate  non-Hermitian Systems }
\author{ Amin Jazaeri and  Indubala I. Satija\cite{email}}
\address{
 Department of Physics and \\
 School of Computational Sciences \\
 George Mason University,\\
 Fairfax, VA 22030}
\date{\today}
\maketitle
\begin{abstract}
A class of one-dimensional lattice  models with incommensurate  complex 
potential $V(\theta)=2[\lambda_r cos(\theta)+i \lambda_i sin(\theta)]$
is found to exhibit localization transition at $|\lambda_r|+|\lambda_i|=1$.
This transition from extended to localized states manifests in the behavior 
of the complex eigenspectum. In the extended phase, states with real 
eigenenergies have finite measure and this measure goes to zero in the 
localized phase. Furthermore, all extended states exhibit real spectrum
provided $|\lambda_r| \ge |\lambda_i|$. Another novel feature of the system 
is the fact that the imaginary part of the spectrum is sensitive to the 
boundary conditions {\it only at the onset to localization}.
\end{abstract}

\narrowtext

Incommensurate systems such as the Harper equation\cite{Harper} provide an 
important class of models exhibiting both extended and localized states 
in one dimension (1D). In this paper, we study the localization transition 
in systems with competing length scales where the underlying potential is 
complex. The systems under investigation are described by the following 
class of lattice tight binding models (tbm),
\begin{equation}
\psi_{n+1}+\psi_{n-1}+2[\lambda_r cos(\theta_n)+ i \lambda_i sin(\theta_n)] 
\psi_n = E \psi_n,
\end{equation}
where $\theta_n = 2 \pi \sigma n+\alpha$. Here, $\alpha$ is a constant phase 
factor and $\sigma$ is an irrational number, which we choose to be the the 
golden mean. This lattice model describes a system where the period of the 
potential is incommensurate with the periodicity of the lattice. 
For $\lambda_i=0$, it reduces to the Harper equation which exhibits a
localization transition at $|\lambda_r|=1$. Recently, non-Hermitian 
systems have been the subject of various 
theoretical\cite{Jones,nelson,sil,Berrycomplex} and 
experimental\cite{aoptics,kudrolli} studies. 
In certain 1D random systems, where all states for the Hermitian problem are 
localized, the addition of a complex vector potential  
has been shown\cite{nelson} to result in the 
delocalization of states, accompanied by the eigenvalues becoming complex.
The system we investigate exhibits both extended and localized states and 
hence facilitates the study of non-Hermiticity on both these phases. 
By investigating how the non-Hermiticity alters the localization transition 
and as well as the eigenenergies in the complex plane, we are able to 
study the correlation between the nature of the eigenspectrum and the 
transport characteristics of the model. 
It is implicit in the literature\cite{ferry} that non-Hermiticity
corresponds to dissipation and decoherence and as such we arguably
explore the impact of these effects on the localization transition. 
Two interesting limits that we explore in detail are the case of pure 
imaginary potential $\lambda_r=0$ and the case where the real and the 
imaginary parts are equal ($\lambda_r=\lambda_i$) which can be 
described as the strong and the weak dissipative limits, respectively.
Further, we also study the case of non-Hermitian lattice models with
real spectra. This is an interesting problem, since certain complex 
potentials in quantum mechanics are known to have a real spectrum 
provided the potentials exhibit parity and time 
reversal(PT) symmetry\cite{Jones}. Here, we seek the criterion for a real 
spectrum in non-Hermitian lattice models exhibiting localization transition.

We study periodic boundary conditions (PBC) anti-periodic boundary 
conditions (APBC) and free boundary conditions to investigate the 
sensitivity to the boundary effects. As expected, only extended states 
are sensitive to the boundary conditions and this sensitivity to the 
boundary effects can be used to distinguish extended and localized states. 
We use $\Delta E_r$, which is 
the real part of difference in the eigenenergies between PBC and APBC, summed 
over all states, to distinguish extended and localized states: The extended
states are characterized by finite value of $\Delta E_r$ while in the localized
phase $\Delta E_r=0$, reflecting its insensitivity to changes at the 
boundaries. Our detailed numerical study, based on sensitivity to various 
boundary conditions and wave functions shows that the non-Hermitian system 
exhibits localization transition at $|\lambda_r| +|\lambda_i| = 1$. This
implies that the Hermitian and the non-Hermitian part of the potential carry
the same weight in determining the transport characteristics of the model.
Fig. (1) shows $\Delta E_r$ and $\Delta E_i$ for weak dissipation limit 
$\lambda_r=\lambda_i=\lambda$, which is discussed in detail below. A rather 
intriguing result is that the imaginary part of the $\Delta E$ is sensitive 
to the boundary effects only at the onset to localization transition. 
This result found to be true for other parameter values implies that
the life time of the metastable system depends upon the boundary
conditions only at the transition point.

In order to explore the relationship between the localization character and 
the behavior of the eigenenergies, we have extensively studied of 
the eigenspectrum in the two-parameter ($\lambda_r, \lambda_i$) space.
Figs. (2-4) show the variation in eigenenergies with the parameters for
some special cases. These studies suggest that it is the localized phase 
of the non-Hermitian lattice model (1) that is characterized by a complex 
spectrum.
 in contrast to previous results\cite{nelson} where the spectra 
becomes complex when the localized states become delocalized. 
In the extended phase, the spectrum is real provided 
$|\lambda_r|\ge|\lambda_i$. Furthermore, for $|\lambda_r| < |\lambda_i|$, 
the number of extended states with real eigenenergies have finite measure.  
Therefore, the fraction of states with real spectrum is finite in extended phase and 
vanishes in localized phase. This provides a new order parameter for the localization 
transition, as shown in Fig. (5).

In contrast with the earlier results\cite{nelson} where a non-Hermitian 
vector potential was found to delocalize the localized states of the 
random system, the addition of a non-Hermitian potential to the Harper 
model which exhibits both extended and localized states does not alter 
the localization character of the system.  Furthermore, our results 
associate a complex spectrum with the localized states, in marked contrast 
with the earlier result where complex spectrum implied delocalization. 
This is one of the central result of our analysis. It implies that
the previous results relating complex eigenvalues and delocalization 
must be understood as special to the type of the system investigated, 
namely random system with non-Hermitian vector potential, and may not 
describe the generic property of non-Hermitian systems exhibiting localization.

We now discuss the two limiting cases: the strong and the weak dissipation 
limit. The first case corresponds to pure imaginary potential($\lambda_r=0$).
It is interesting to note that the model with pure imaginary potential
exhibits duality very similar to that of the pure Hermitian 
problem\cite{aubry}. Under the Fourier transformation (FT),
\begin{equation}
\psi_n = \sum  e^{ i \theta_n m } \phi_m
\end{equation}
the tbm (1) with $\lambda \equiv \lambda_i$ reduces to
\begin{equation}
\phi_{m+1}+\phi_{m-1}+\frac{2i}{(\lambda)} cos(\theta_m) \phi_m = 
-\frac{iE}{\lambda} \phi_m.
\end{equation}
Comparing this with the original model, we obtain that
\begin{equation}
E^*(\lambda) = -\frac{i}{\lambda} E (\frac{1}{\lambda}).
\end{equation}
This implies that the real and the imaginary part of the eigenenergies
are related as $E_r (\lambda) = \frac{1}{\lambda} E_i(\frac{1}{\lambda})$. 
Therefore the case of the pure imaginary potential has the interesting 
property that the localization transition interchanges the real and the 
imaginary part of the spectrum. At the onset to localization ($\lambda=1$) 
the model is self-dual, with $E_r=E_i$. Fig. (6) shows eigenenergies at 
some values of the parameter in the extended phase. Due to duality these figures 
also show the spectrum in the localized phase with the interchange 
of $E_r$ and $E_i$. At the onset of localization, the spectrum with 
$E_r=E_i$ resembles that of the Harper equation. It is intriguing that 
even in this strong dissipative limit, extended states with real spectrum 
have a finite measure. In this lattice model with competing length scales, 
it therefore appears that extended states are essential for obtaining real 
eigenenergies. This is to be compared to earlier formal results where 
PT symmetry was a key for obtaining real spectrum for complex potentials.

We next discuss the weak dissipation limit, 
$\lambda_r=\lambda_i \equiv \lambda$, described by the following tbm,
\begin{equation}
\psi_{n+1}+\psi_{n-1}+2\lambda e^{i\theta_n}\psi_n = E \psi_n
\end{equation}
As shown in figure (1), the system exhibits localization transition
at $\lambda=.5$. The localization threshold is half of that of the Harper equation because both
the real and the the imaginary part of the potential contributes towards localization.
In the extended phase, the spectrum is found to be real and identical to that of $\lambda=0$ limit,
namely $E=2cos(\theta_n)$. This explains the constant value of $\Delta E_r$ in the extended phase
as seen in figure (1). In the localized phase , the eigenenergies are complex
and appear to be described by the following expression,
\begin{eqnarray}
E_r &=& 2cosh(\gamma) cos(\theta_n)\\
E_i&=& 2sinh(\gamma) sin(\theta_n).
\end{eqnarray}
Here $\gamma$ is the inverse localization length of the system which is
found to be equal to the corresponding value for the Harper equation
$\gamma=log(2\lambda)$. In the limit $\lambda \rightarrow \infty$, 
eigenenergies lie on a circle. Therefore, the localized phase is 
metastable with the lifetime determined by the localization length.

The FT of the model (5) can be analysed further as the transformation  (2)
reduces tridiagonal matrix to the following triangular matrix: 
\begin{equation}
2 cos(\theta_m) \phi_m + 2 \lambda \phi_{m-1} = E \phi_m.
\end{equation}
With PBC, the eigenvalues of this triangular marix are the solution of 
the following algebraic equation:
\begin{equation}
(2\lambda)^N = \prod(E-2cos(\theta_n))
\end{equation}
For $\lambda < 0.5$, as $N \rightarrow \infty$, we obtain $E=2cos(\theta_n)$,
which is found to be identical with  numerically obtained spectrum of the 
model(5). For $\lambda > .5$, the FT of the model allows real solutions: 
$E= 2cos(\theta_n)+2\lambda$. These real energies were not found to be the 
solutions of the model (5). The FT of the  model also exhibits complex 
solutions. It is easy to see that in $\lambda \rightarrow \infty$ limit,
the algebric equation (9) has a solution where $E/(2\lambda)$ lies on 
the unit circle which is also the solution for the model (5).
This indicates a deep relationship between the spectra of the model (5) 
and its FT -- the details have proven so far elusive. 

Another aspect of model (5) is that at the onset to localization 
$\lambda=0.5$, the FT of the model describes the strong coupling limit 
of the fluctuations of the Harper equation once the exponentially decaying 
part is factored out\cite{KSprl}. This also describes the Ising model 
at the onset to long range order for $E=0$\cite{Lieb}. This limit has been 
shown to be universal using renormalization methods\cite{KSprl} as well 
as more rigorous analytic tools\cite{andy}. This result therefore 
establishes the multi-fractal character of the FT of the wave function at 
the onset of the localization transition. It should be noted 
that the eigenspectrum remains continuous at the localization transition 
in contrast with the Harper equation which is characterized by 
singular-continuous spectrum at the transition. Therefore,
the $\lambda_r=\lambda_i$ limit of the model (1) provides a new class of 
incommensurate systems where the eigenspectrum is gapless and remains 
continuous except in the localized phase.

In summary, the localization transition of the Harper equation remains 
unaffected by the non-Hermitian perturbation. Weakly dissipative system 
are characterized by real eigenenergies in the extended phase. As the 
strength of the non-Hermitian potential increases, the number of extended 
states with real eigenenergies decrease approaching zero at the onset 
to localization. In the localized phase, states with complex spectrum have 
full measure. The localized phase is metastable, with state lifetimes 
determined by the localization length. The question of real eigenvalues is 
determined by 
both the transport character of the states as well as the amount of 
dissipation. An interesting result is that the extended states with real 
eigenenergies always have a positive measure while localized states with 
real energies have zero measure. Therefore, the measure of real eigenenergies 
provides a new order parameter for localization-delocalization transition.

In contrast to the earlier results on localization in non-Hermitian 
systems\cite{nelson}, the system studied here associates a complex spectrum 
with the localized phase. Further, the non-Hermitian potential does not 
delocalize the localized states.The localized phase of the Harper equation 
remains localized for any amount of non-Hermitian perturbation. This may be 
related with the fact that the non-Hermiticity appears in the diagonal part 
of the potential while the the non-Hermitian vector potential studied 
earlier\cite{nelson} affects the off-diagonal part of the lattice model.

One future aspect of study is the extension of these results to a
classical non-integrable (perhaps a kicked) system, thus investigating
the effect of dissipation on non-integrable systems exhibiting localization.
Furthermore, it should be noted that the equation (1) is the fermion 
representation of isotropic XY spin-$1/2$ chain in a complex magnetic 
field which is spatially modulating\cite{Lieb}. The consequences of the 
localization transition with complex spectrum on the magnetic properties 
of the system is another interesting open question.
Finally, localization transition in incommensurate tightbinding lattice models corresponds to
a transition to strange nonchatoic attractors (SNA) \cite{SNA} in quasiperiodically driven maps.
Therefore, the results of this paper may have important implications in the study of SNA's
in complex maps.

The research of IIS is supported by a grant from National Science
Foundation DMR~093296.
We would like to thank Arjendu Pattanayak for useful discussions.

\begin{figure}
\caption{ (a) and (b) respectively show $\Delta E_r$ and  $\Delta E_i$
vs $\lambda \equiv \lambda_r=\lambda_i$ for $\sigma = 377/610$. The
$\Delta E_i$ appears to be related to
the derivative of $\Delta E_r$.}
\label{fig1}
\end{figure}

\begin{figure}
\caption{ (a) and (b) respectively show the variation in eigenvalues as a function of 
$ \lambda \equiv \lambda_r=\lambda_i$ for PBC for $\sigma=34/55$. The extended phase of this non-Hermitian system
exhibits real spectrum.
Note that in contrast to the Harper equation, there is a bending and merging of levels at the transition.
Furthermore, unlike Harper equation, the spectrum is not symmetric about $\lambda=0$.}
\label{fig2}
\end{figure}

\begin{figure}
\caption{ (a) and (b) respectively show the variation in eigenvalues as a function of $\lambda \equiv \lambda_r$ for fixed
$\lambda_i=.25$ for  PBC for $\sigma=34/55$. In the extended phase all states have real eigenenergies provided 
$\lambda_r \ge \lambda_i$.   }
\label{fig3}
\end{figure}

\begin{figure}
\caption { Same as figure 2 and 3 for pure imaginary potential $\lambda_r=0$. }
\label{fig4}
\end{figure}

\begin{figure}
\caption{shows the fraction of states with real eigenenergies as a function of $\lambda$ for $\lambda_i=.25$ case (a)
and $\lambda_r=0$ (b). Here $\sigma=233/377$. The steps seen in this plot are due to finite size effects}
\label{fig5}
\end{figure}

\begin{figure}
\caption{ Figure shows the spectrum for differrent values of $\lambda$ for $\sigma = 55/89$.
For pure imaginary potential. Three verical columns respectively correspond to $\lambda=.5,.75,1$:
The three rows show
(a) $E_r$ vs $E_i$ (b) $E_r$ vs n (c) $E_i$ vs n(sorted independently of the real part)}
\label{fig6}
\end{figure}

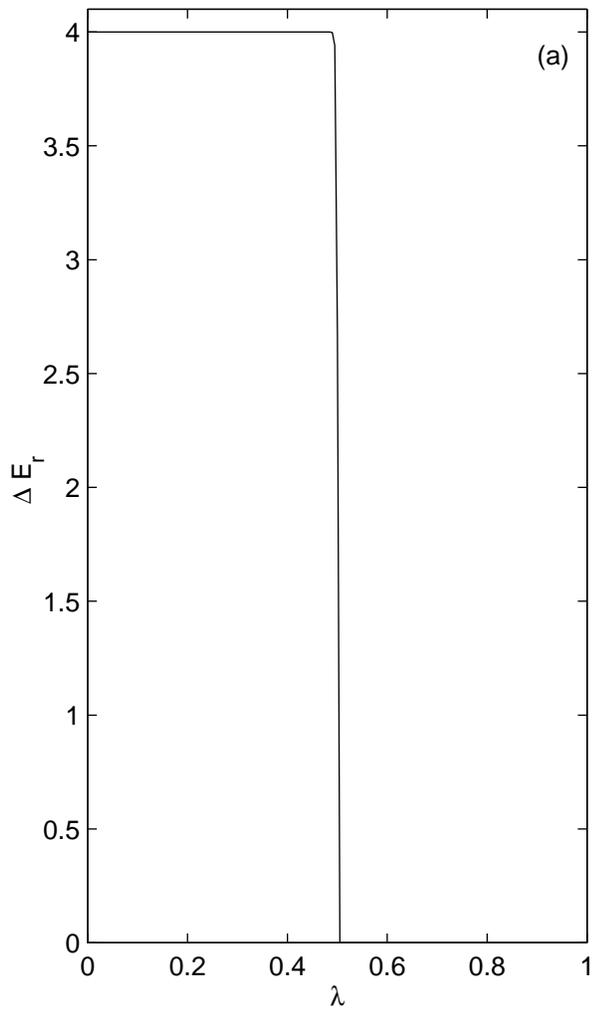
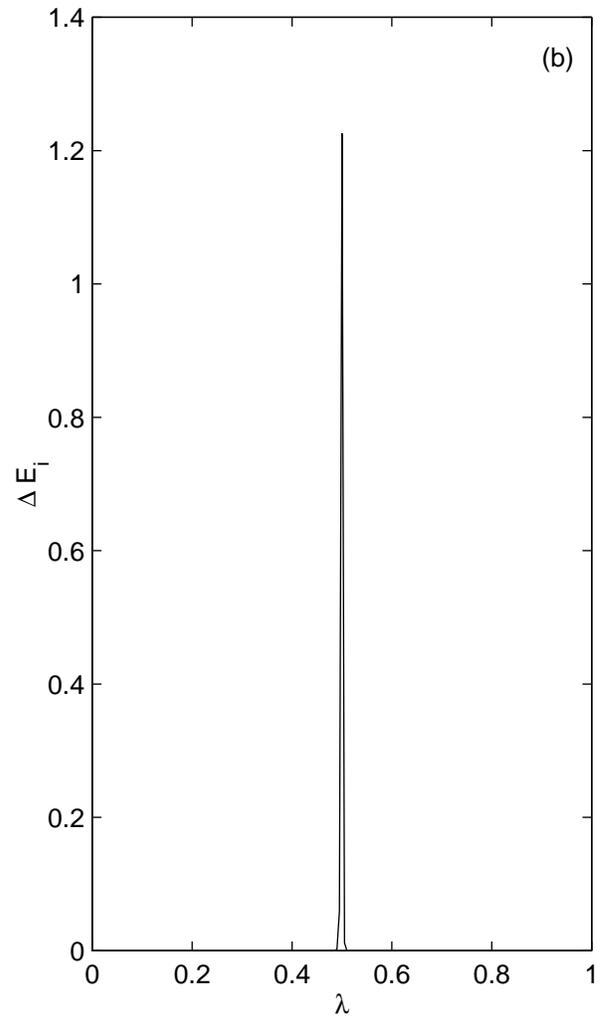
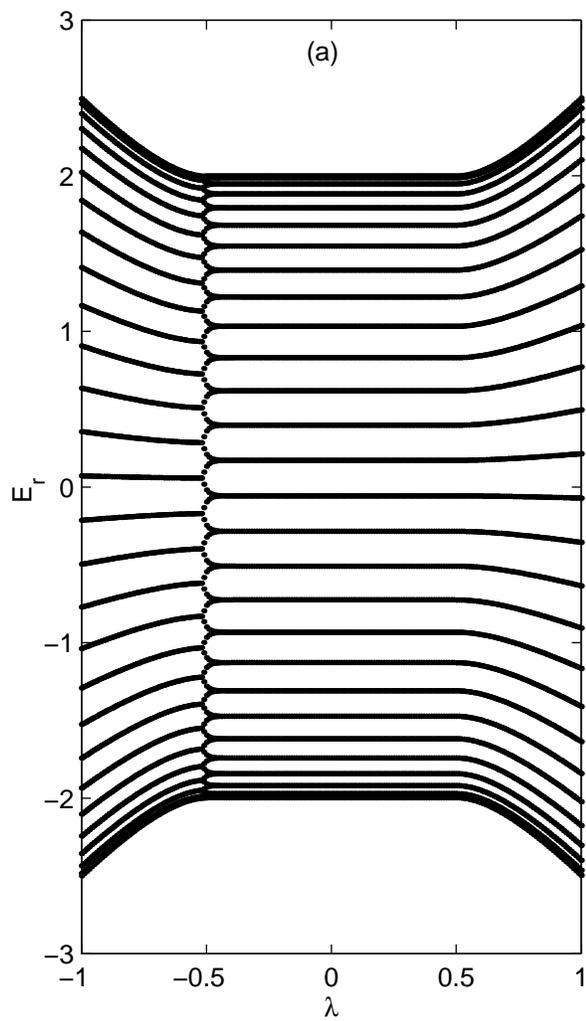
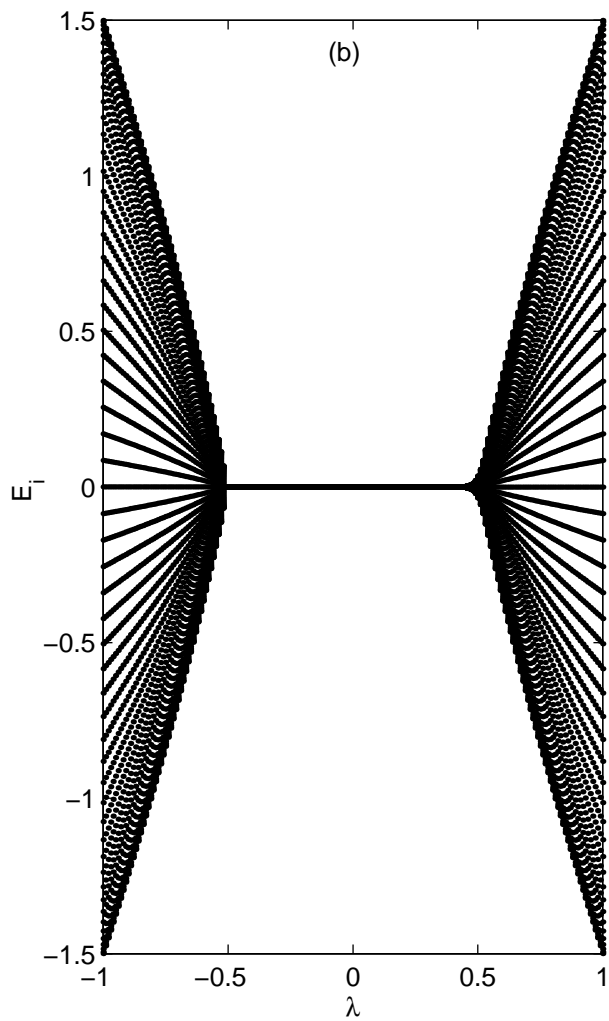
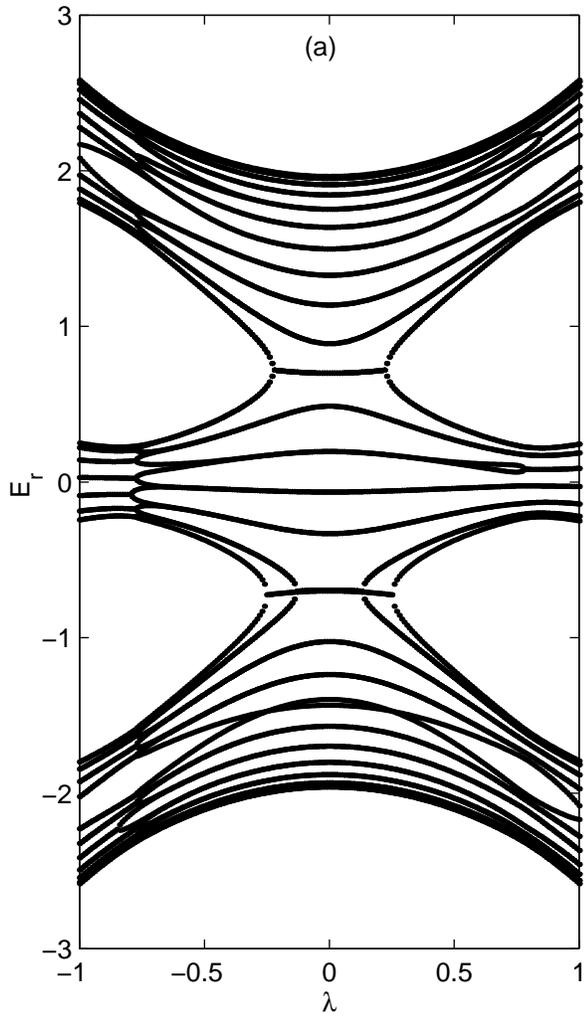
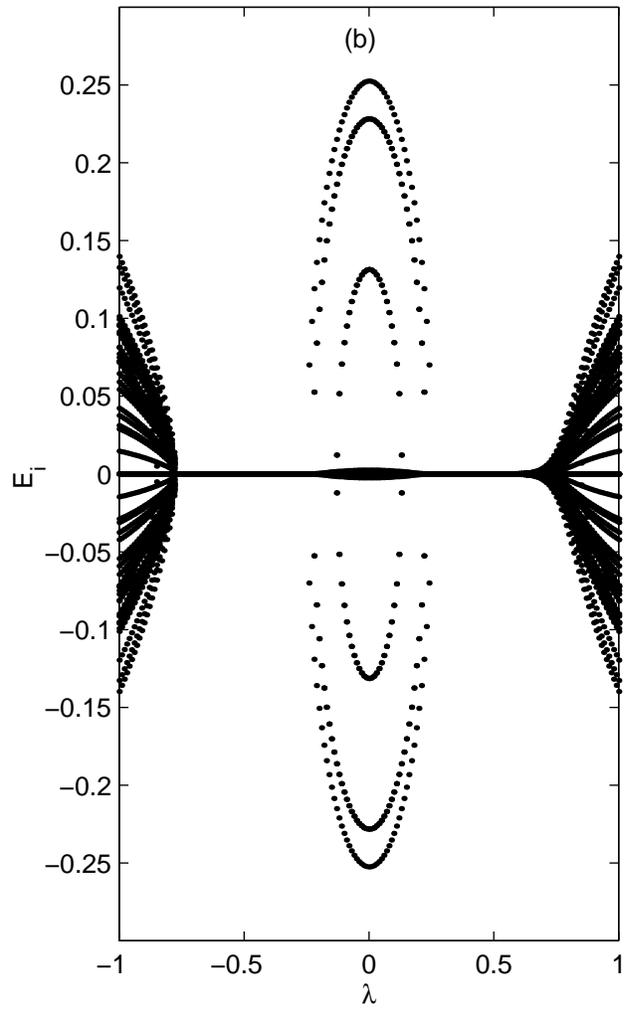
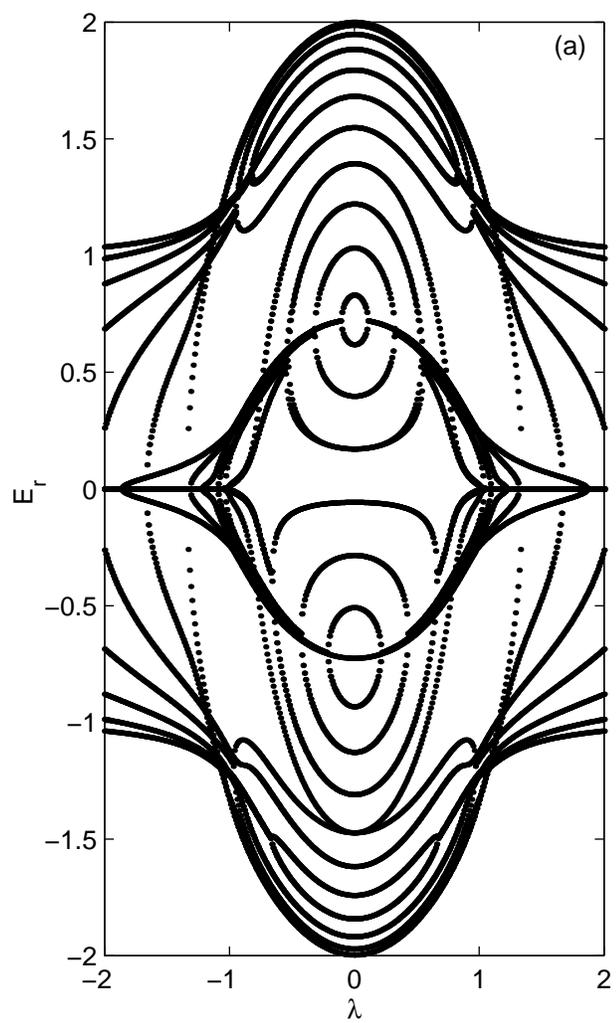
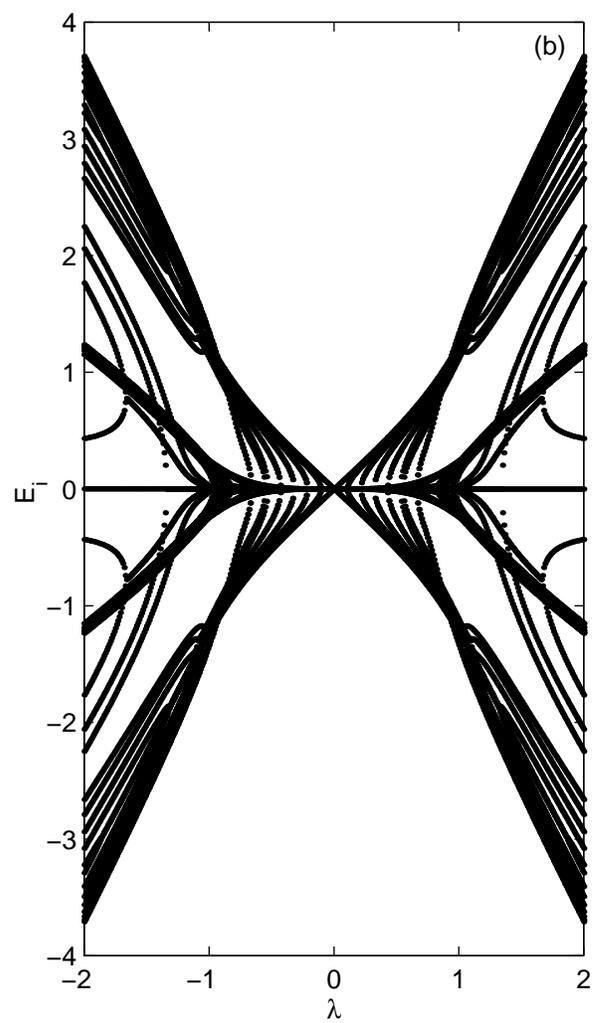
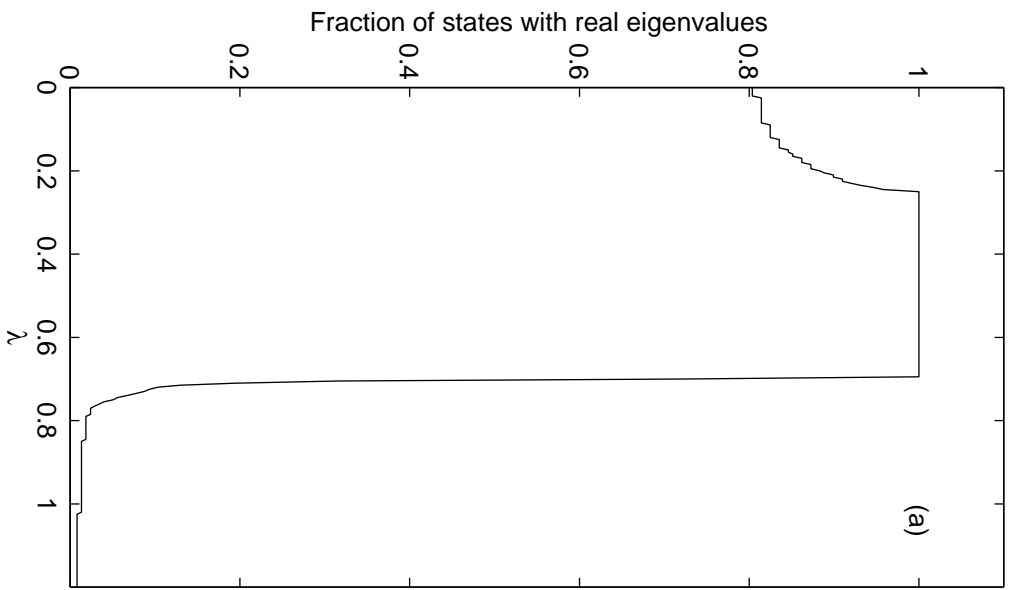
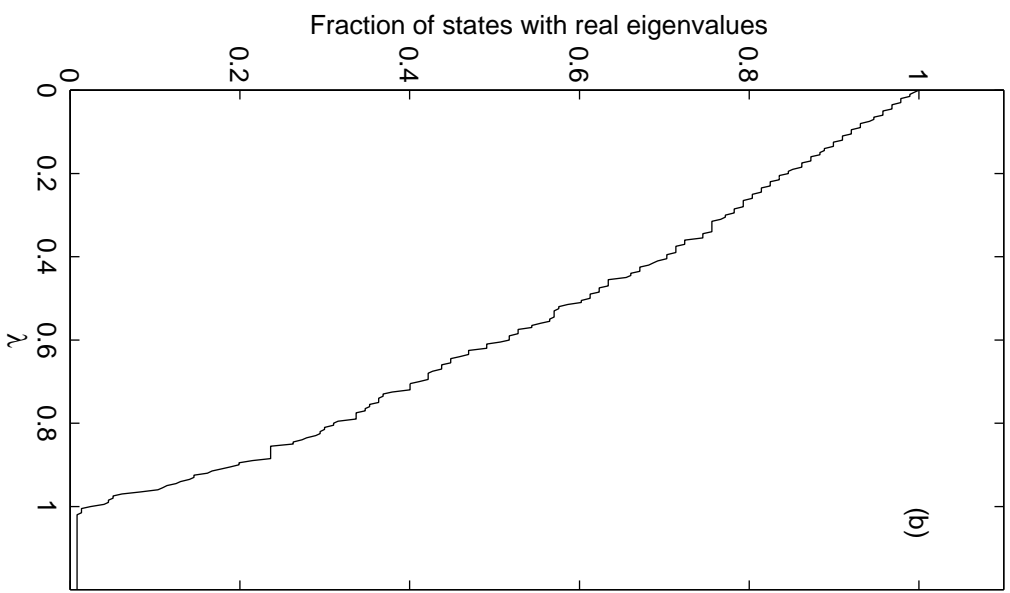
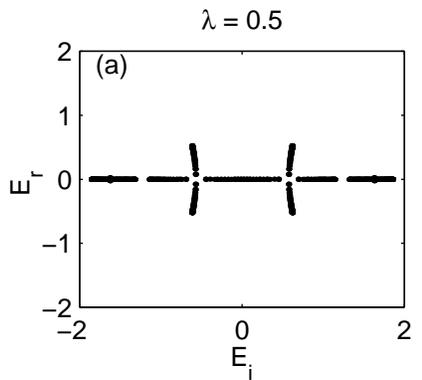
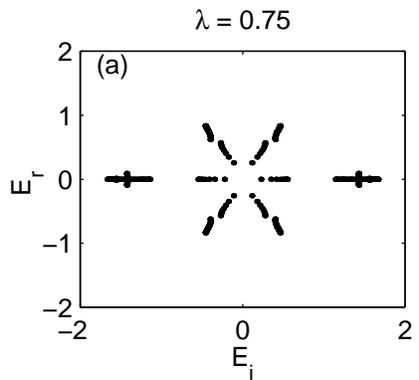
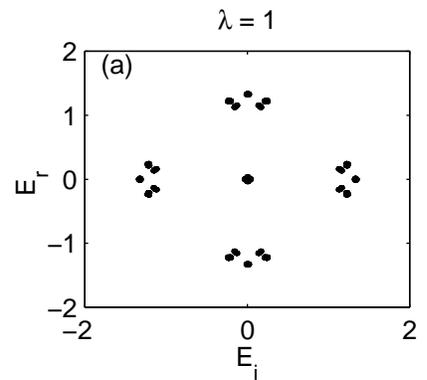
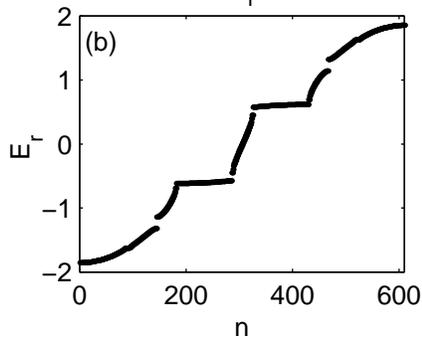
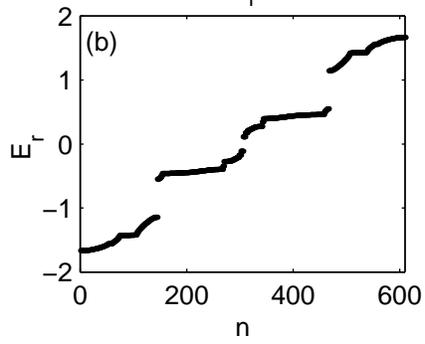
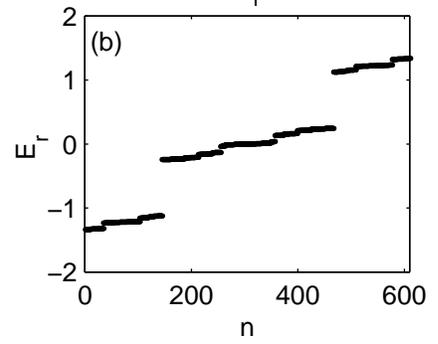
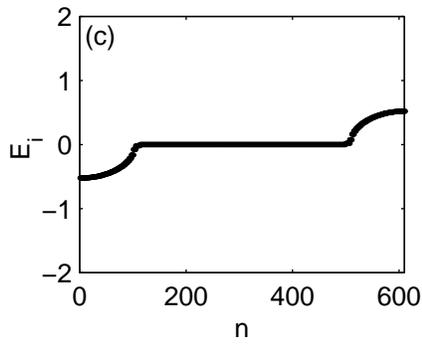
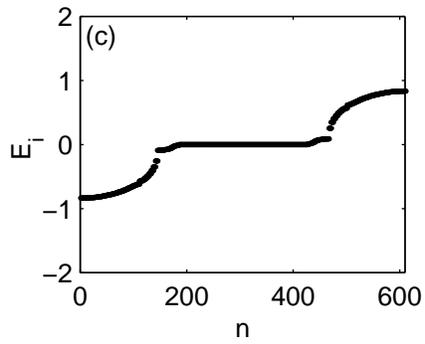
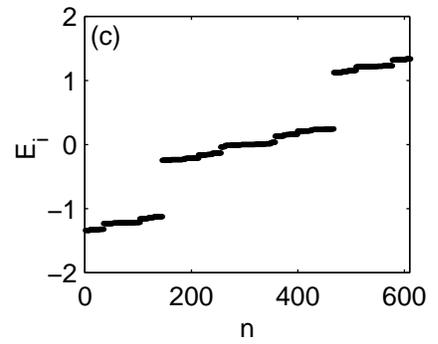
\end{document}